# The Science of Urban Metabolism and Sustainability

Mariana Brüning-González[1,2,3], Jose Ignacio Arroyo[4,5,6], Pablo A. Marquet[4,5,7,8,9], Horacio Samaniego[2,6*]

1 Programa de Doctorado en Ciencias mención Ecología y Evolución, Escuela de Graduados, Facultad de Ciencias, Universidad Austral de Chile, Valdivia, Chile.
2 Laboratorio de Ecoinformática, Instituto de Conservación, Biodiversidad y Territorio, Universidad Austral de Chile, Campus Isla Teja, Valdivia, Chile.
3 Instituto de Ciencias Ambientales y Evolutivas, Facultad de Ciencias, Campus Isla Teja, Universidad Austral de Chile, Valdivia, Chile.
4 Santa Fe Institute, Santa Fe, NM 87501, USA.
5 Center for Mathematical Modeling, Santiago, Chile
6 Department of Environmental Studies, New York University.
7 Facultad de Ciencias Biológicas, P. Universidad Católica de Chile, Alameda 340, Santiago, Chile.
8 Instituto de Sistemas Complejos de Valparaíso, Subida Artillería 470, Valparaíso, Chile.
9 Centro de Cambio Global UC.
* Corresponding Author (horacio@ecoinformaica.cl)

## Abstract

Understanding the quantitative patterns behind scientific disciplines is fundamental for informed research policy. While many fields have been studied from this perspective, Urban Science (USc) and its subfields remain underexplored. As organisms, urban systems rely on materials and energy inputs and transformation (i.e. metabolism) to sustain essential dynamics. This concept has been adopted by various disciplines, including architecture and sociology, and by those focused on metabolic processes, such as ecology and industrial ecology. This study addresses the structure and evolution of Urban Metabolism (UM) and Sustainability research, analyzing articles by disciplines, study subjects (e.g., cities, regions), methodologies, and author diversity (nationality and gender). Our review suggests that UM is an emerging field that grew until 2019, primarily addressed by environmental science and ecology. Common methods include Ecological Network Analysis, and Life Cycle Assessment, and Material Flow Analysis, focusing flows over stocks, ecosystem dynamics and evolutionary perspectives of the urban system. Authors are predominantly from China and the USA, and there are less gender gaps compared to general science research. Our analysis identifies relevant challenges that have become evident in the statistical properties of this scientific field and which might be helpful for the design of improved research policies.

**Highlights:**

- Urban Metabolism (UM) is an emerging interdisciplinary field, primarily focused on material and energy flow quantification, with limited attention to structural and systemic aspects.
- UM studies adopt diverse sustainability definitions, from resource and waste efficiency to approaches integrating ecosystem resilience and human well-being.
- Limited integration of scaling and stocks-based approaches in UM research highlights opportunities to address dynamic, systemic, and global urban sustainability challenges.
- Urban Metabolism and Sustainability research has experienced a decline as of 2019; has been mainly focused on small spatial scales, and has been conducted mostly by male researchers in the USA and China.

# 1. Introduction

Understanding the quantitative patterns associated with the structure and evolution of scientific disciplines has become relevant not just to better grasp the emergent challenges in a given field, but also to make informed policy decisions associated with, for example, research funding (Fortunato *et al.*, 2018; Zeng *et al.*, 2017). In general, science has been studied under this perspective in different specific fields. One of the fields that remains understudied in this perspective is urban science, particularly the intersection between urban metabolism and sustainability.

In 2007, urban populations exceeded rural inhabitants, with over 4 billion city dwellers by 2023, accounting for 57% of the global population (UN-Habitat, 2022; World Bank, 2023). This rapid urbanization has led to significant environmental challenges, including land-use changes, habitat fragmentation, invasive species introduction, and resource depletion (Alberti, 2024; Bai *et al.*, 2017; Elmqvist, 2013; Loreau *et al.*, 2022). However, cities are also the framework for innovations and problem solving, presenting new opportunities for sustainable development. This dual nature makes cities a key focus in the Anthropocene (Seto *et al.*, 2017).

Urban science (USc) is a multidisciplinary human behavior science addressing the dynamic interplay between human activities and knowledge exchange related to urban development. It seeks to integrate urban, scientific, and social aspects into a body of knowledge attempting to describe and understand the dynamics of one of the ultimate constructs of humanity: the city (Lobo *et al.*, 2020). USc has emerged as a pivotal discipline in the integration and synthesis of planning across several realms, from scientific and civil organizations, to firms and governments (Bettencourt *et al.*, 2020). This reinforced the prevalent paradigms of cities as complex systems. For example, the geographer Michael Batty reinforced the longstanding idea that cities are "complex systems that mainly grow from the bottom up, their size and shape following well-defined scaling laws that result from intense competition for space" (Batty, 2008). Since a system is composed by a set of interconnected elements, a city may be viewed as a complex system made up by urbanites, firms, and social structures organized in a hierarchical decision-making process modulating flows of matter, information, and energy around a shared ecology of interconnected elements, just like as ecosystems, communities, and populations do. This provides a more general theoretical foundation for USc, anchored on the analogy of cities as organisms (Bettencourt, 2013; Samaniego & Moses, 2008).

This parallelism shows analogous characteristics to the fundamental biological process known as metabolism that encompasses the transformation of matter procured from the environment into usable energy and waste. A historical analysis reveals that diverse approaches are used to elucidate the complexities of metabolism within biological systems (see citations in Supplementary Table 1). Already in 1614, Santorio Santorio conducted the first documented investigation into human metabolism, demonstrating a quantifiable and predictable relationship between ingested food mass and excreted fecal matter (Céspedes Restrepo & Morales-Pinzón, 2018). Later on, Kleiber established that the metabolic rate in mammals scales as a power law of the animal's body mass with an ¾ exponent (Kleiber, 1932). This discovery has sparked strong scientific inquiries regarding the underlying mechanisms responsible for such a predictable behavior (Bettencourt *et al.*, 2007; West *et al.*, 1997). So far, it has spurred the development of rigorous research efforts aimed at understanding how intrinsic optimization processes shape an organism's fitness (Firn & Jones, 2009). This is partly due to the pivotal role in which metabolism dictates the rates of all biological processes, including growth, reproduction, and evolution.

The term metabolism has also been used in socio-technological systems, such as cities coining the term: Urban Metabolism (UM). Already in the seventeenth century, William Harvey drew the analogy between urban transportation systems and blood movement in the circulatory system. Later, in 1859, Karl Marx used the term *stoffwechsel* (i.e. metabolism) to refer to the cyclical interdependence of man and nature in urban systems. All these elements have concurred to the current definition of UM proposed by Kennedy, claimed as "the total sum of the technical and socioeconomic processes that occur in cities, resulting in growth, production of energy, and elimination of waste" (Burkett & Foster, 2006; Céspedes Restrepo & Morales-Pinzón, 2018; Kennedy *et al.*, 2011; Saito, 2017; Wolman, 1965).

Recent theoretical advancements in Urban Metabolism (UM) have incorporated the Metabolic Theory of Ecology (MTE), which posits metabolic rates as a fundamental driver of ecological processes across scales (Brown *et al.*, 2014). This perspective aligns with analogies drawn between cities and biological systems, such as the human immune system or cellular structures (Bristow & Mohareb, 2020; Chang *et al.*, 2021). While some critics question the ecological theoretical grounding of UM (Golubiewski, 2012), scholars like West have contributed significantly to the understanding of scaling laws in both natural and anthropogenic systems, offering insights into urban dynamics, behavior, and structure (West, 2017).

Metabolic theory is also applied to a city in the Urban Scaling Theory (UST) (Bettencourt *et al.*, 2007). In cities, as in organisms, many quantities scale with the total population, or cells, according to a scaling law of the form (Equation 1):

$$Y = Y_0 \cdot N^a \qquad (1)$$

Where Y is some city rate, time, or quantity, such as Gross Domestic Product (GDP), water consumption, or urban area; N is the number of individuals in the city; $Y_0$ is a constant; and $a$ is an exponent.

This exponent typically exhibits a range between 0 and 2. Values below 1 indicate a sublinear relationship, suggesting that as the city's size doubles, its quantity increases by less than double. This economy of scale implies that growth in size results in reduced size-related benefits or a decrease in per capita costs associated with increased size. Conversely, exponents greater than 1 indicate a superlinear relationship, wherein the increase in quantity surpasses the rate of size expansion. This superlinearity can be attributed to the facilitation of human interactions within urban networks, which leads to a disproportionate increase in urban productivity as city size grows (Cheng, 2023). However, it is crucial to acknowledge that increases in size may also bring about negative consequences, such as higher crime rates, elevated pollution levels, and increased stress. To clarify these scaling relationships, various modeling frameworks have been developed (Ribeiro & Rybski, 2023). Other models analogous to the organism's growth are shown in Box 1.

**Box 1 City Growth Analogies**

Briefly, in West's model, a city's growth is a simple model analogous to the growth of organisms:

$$\frac{dN}{dt} = a \cdot N^b - c \cdot N \qquad (2)$$

Where $a$, $b$, and $c$ are reduced parameters. The solution of this equation is:

$$N = (h + k \cdot \exp(-l \cdot t))^m \quad (3)$$

This model describes the logistic growth of the population in cities, and *h*, *k*, *l*, and m are reduced parameters. Under a given set of parameters, it even predicts collapse, which evidences the challenge of moving the observed set of parameters to a sustainable scenario (Bettencourt, 2013).

Another relevant constraint of metabolism in biological systems is temperature. The temperature response of metabolism and most biological processes are characterized by an initial exponential increase (often modeled using the Arrhenius equation) followed by an abrupt and more rapid decay after a given critical point. Briefly, the Arrhenius model is defined by the following equation:

$$Y = Y_0 \cdot \exp(-b/T) \quad (4)$$

Where T is the temperature in Kelvin. Quantities such as time, mortality rate, or metabolic rate in endotherms follow an inverse –convex– pattern where the variables increase at low or high temperatures. Cities and other social and economic systems also exhibit a temperature dependence as many processes are associated with the kinetic energy input. For example, world economic production follows a concave temperature dependence, and energy uses a convex. These curved patterns can be modeled using different unimodal models going from empirical to first-principles-based approaches (see, for example, Arroyo *et al.*, 2022).

The analogy between cities and biological organisms is compelling. By examining materials and energy flows within urban systems, researchers aim at understanding and optimizing urban processes for sustainability. Studies focusing on these flows (Haberl *et al.*, 2019; Lin & Grimm, 2015; Lucertini & Musco, 2020; Wang *et al.*, 2011) offer valuable insights. However, a key question remains: What constitutes a city's metabolism, and how does it connect to sustainable development?

## 1.1. Urban Metabolism and Sustainability

While the term 'sustainability' may be contested in light of the global climate and ecological crisis, a widely accepted definition emphasizes meeting current needs without compromising the ability of future generations or the biosphere's capacity to absorb human impacts (Brundtland, 1987). Cities are a critical focus area for achieving sustainability (Bai *et al.*, 2022, 2024; Newman & Jennings, 2008), as evidenced by the UN Sustainable Development Goal 11: "Sustainable Cities and Communities" (https://sdgs.un.org/goals/goal11). UM research offers a valuable tool for measuring urban sustainability by analyzing the resource flows that sustain city functions. Through resource efficiency, most UM research aims to minimize the environmental impacts of human activity within cities, such as resource extraction from ecosystems and waste generation (Elmqvist, 2013; Rees & Wackernagel, 1996).

While UM studies have predominantly focused on material and energy flows, they have enabled explicit analyses of urban system sustainability (Chen & Chen, 2014; Cui, 2018; Newman, 1999). We emphasize the sustainability of the urban system rather than sustainability within urban spaces, acknowledging cities as complex systems with impacts extending beyond their geopolitical boundaries. This perspective aligns with an urban ecology approach, rather than an ecology-in-cities approach (Hall & Balogh, 2019). For instance,

Huang *et al.* (2015) argue that a sustainable city maintains a critical level of local ecosystem function and biodiversity to ensure ecosystem services for human well-being, given the environmental and sociopolitical risks associated with resource importation. A comprehensive understanding of the social-ecological dynamics of urban systems is essential for elucidating their contribution to global sustainability (Lobo *et al.*, 2020).

This work reviews critical aspects of UM. We examine existing methodological approaches and explore the potential of framing cities as ecological systems, extending beyond traditional metabolic flows (e.g., Lobo *et al.*, 2020; Zhang, 2013; Zhang *et al.*, 2015; Backhouse, 2020; Huang *et al.*, 2015; Michalina *et al.*, 2021; Mori & Christodoulou, 2012). We specifically investigate whether UM can be understood from an ecological systems perspective, considering factors beyond matter and energy exchange, while discussing the role of allometric scaling in informing sustainability.

We seek to characterize the existing body of knowledge, identify the primary research foci, classify studies based on their scale of analysis, and determine the main methodologies used. Furthermore, the research aims at discussing the relationship between UM and sustainability, drawing on the most recent scholarly work to provide a comprehensive understanding of this critical urban issue.

# 2. Methods

We assessed how the concept of UM is represented across various disciplines within the literature. We have therefore classified contributions by year, country, and academic discipline based on the categories found in the Web of Science (WoS). This classification aims to identify the novelty of the data employed to support their conclusions.

We retrieved all articles in the WoS that use the terms "Urban Metabolism" and "Sustainability" in the main search, including a search across all fields (i.e., title, abstract, topic) of articles in every issue of the WoS Core collection (WoS, 2024).

The initial assessment showed that foundational articles in the field are not consistently included in the search, which prompted us to additionally include a limited number of handpicked and emblematic articles having a large citation volume in this analysis (see Supplementary Table 2 for a list of articles). To gain further insights into the approaches used in UM research, we examined years of publications, authors' nationalities, research areas according to the WoS classification, and gender disparities. We used the Python library gender-guesser to establish the authors' genders (Arcos & Saeta Pérez, 2016).

The articles were screened following the PRISMA methodology (Preferred Reporting Items for Systematic Reviews and Meta-Analyses) (Moher, 2009). Duplicates and articles deemed not relevant were also removed following Page *et al.* (2021). The screening process to exclude articles not relevant to our purposes consisted in: i) articles describing specific flows of water, energy, or certain chemicals and materials, searching for words like "pollution", "pollutants", and "waste"; ii) articles describing specific case studies, often limited to a particular location deprived of broader impact and potential generalizations, as opposed to contributions that, given their methodological or conceptual innovation, would convey broader generalizations. We assumed that these contributions failed to provide crucial information as they usually describe systems where the economy, consumption behaviors, and the environment are hardly comparable, and; iii) articles published before 2018 with less than 10 citations.

#### 2.1.1. Spatial Scale

The spatial scale of each article was labeled under two types of scales: i) social, i.e. socio-technological or territorial administrative divisions, and; ii) nature, i.e. an ecosystemic or natural division. Social labels include: Country, Region, City, Neighborhood, and Other (e.g., Industry, University Campus). Nature labels include: Planetary, Ecosystem, Basin, and Multi-scale (e.g., articles that show results at global, national, and urban scales). This classification was made according to the article's study object, i.e., if they named cities (e.g., Beijing, Toronto, or Santiago), the geographic scale would be listed as "City". The broader "Planetary" scale, included articles discussing socio-ecological dynamics under a global perspective.

#### 2.1.2. Methodologies in UM

To systematically discuss the methodologies used in UM studies, we initially screened articles based on the presence of keywords in their titles. For articles lacking explicit keyword mentions, we conducted a detailed examination of their Materials and Methods sections.

From an initial pool of 21 methodologies, we identified and summarized the 10 most relevant. These methods were categorized as either accounting or modeling, following Zhang (2015). Accounting methods quantify material and non-material flows, while modeling methods simulate upstream resource consumption. Some methodologies not explicitly categorized by Zhang were included and classified based on their approach. The summary table provides details on each method's goal, description, input data requirements, directionality (top-down or bottom-up), and outputs. A more comprehensive overview of the primary methodologies is presented in Supplementary Table 3.

Finally, we try to elucidate the relation between scales of analysis in the articles reviewed and the methodology used.

#### 2.1.3. Sustainability

To investigate the relationship between UM and Sustainability, we initially identified articles containing the term "Urban Metabolism" in the WoS database. Subsequently, we analyzed these articles' keywords and constructed a bibliometric network using VOSviewer (Van Eck & Waltman, 2010). The network visualized the most frequent keywords as nodes, with connections representing co-occurrence in articles.

We additionally summarized how key works on UM and socio-ecological fields define sustainability, and how UM contributes to achieving this way of understanding sustainability.

# 3. Results

Our WoS search using the terms "Urban Metabolism" and "Sustainability" returned 792 articles. We studied the dynamics of the publications and key concepts, classified according to discipline, spatial scale and methodologies, and analyzed diversity of authors.

## 3.1. Dynamics of the Science of Urban Metabolism and Sustainability

From the dataset's initial year we observe an exponential increase in the number of publications, characterized by a sharp rise in the number of articles related to UM as of 2010.

In fact, 94% of all articles have been published since then (Figure 1). However, since 2019 the number of papers has decreased and the trend has been sustained until 2024.

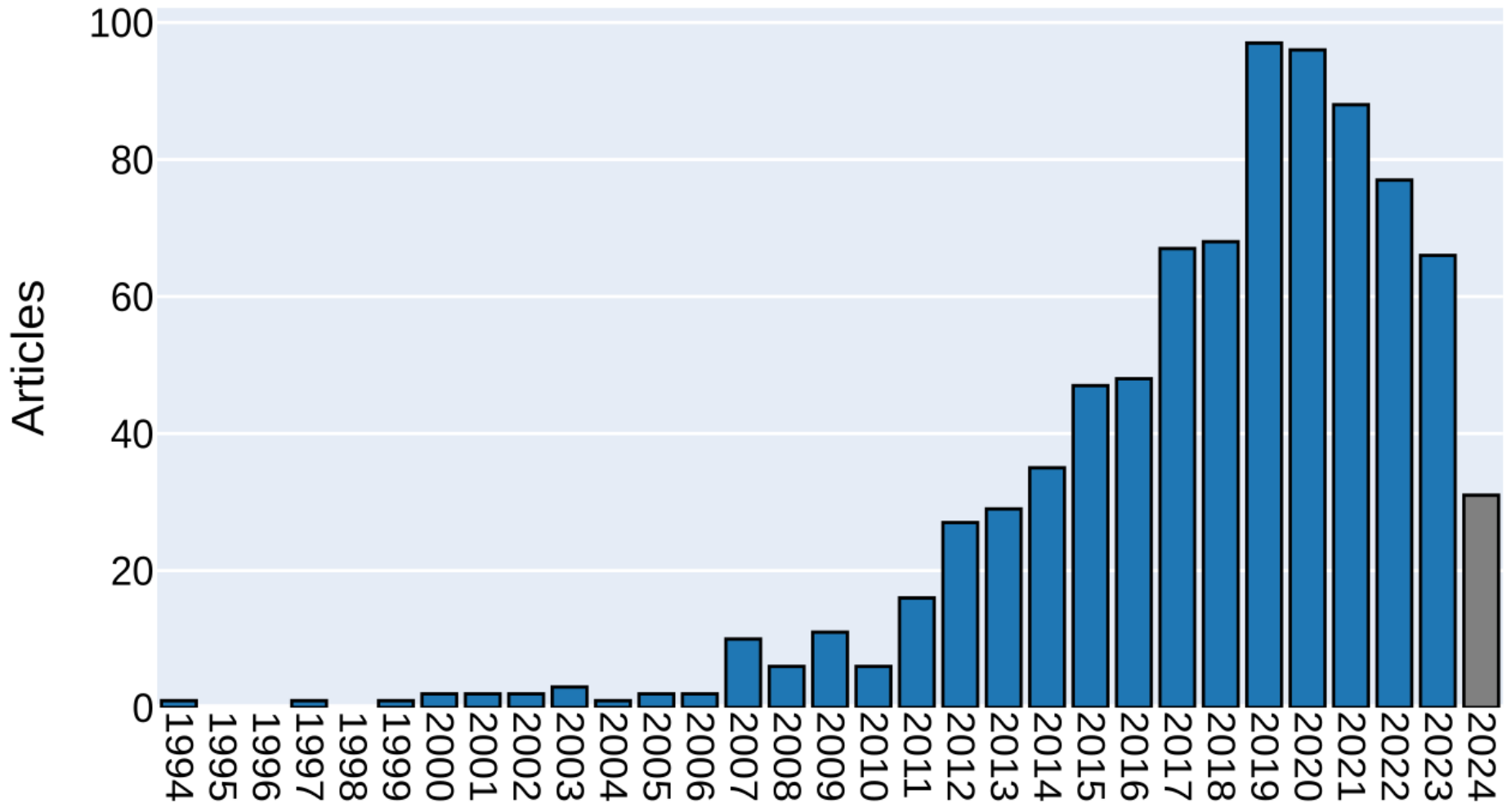

*Figure 1: Number of publications regarding UM and Sustainability in the WoS database through the years, from 1999 to 2024.*

## 3.2. Key Concepts in UM and Sustainability Field

Figure 2 represents a network of keywords in the WoS database using "Urban Metabolism" as search query. The network keywords visualization shows that "Urban Metabolism" is strongly related to "Sustainability" (the third keyword with more occurrences after "Energy" and "Cities"). In addition, also highly used keywords like "energy", "consumption", "water", and "emissions" are concepts associated with resource consumption.

Table 1 summarizes different ways of understanding sustainability in different UM works and how this contributes to sustainable planning. Approaches to sustainability are generally associated with resource use and waste emissions (or their reduction). Similarly, durability over time seems to be a relevant characteristic. Some authors address sustainability at a city level, while others refer to global sustainability or sustainable development. UM's contribution to sustainability mainly depends on quantifying (metabolic) flows and on its use as an input for planning towards sustainability. At the same time, the conception of the city's metabolism, in analogy to a natural system, can help us elucidate how the growth dynamics of cities generates environmental loads and, therefore, provides valuable information for sustainable urban planning policies.

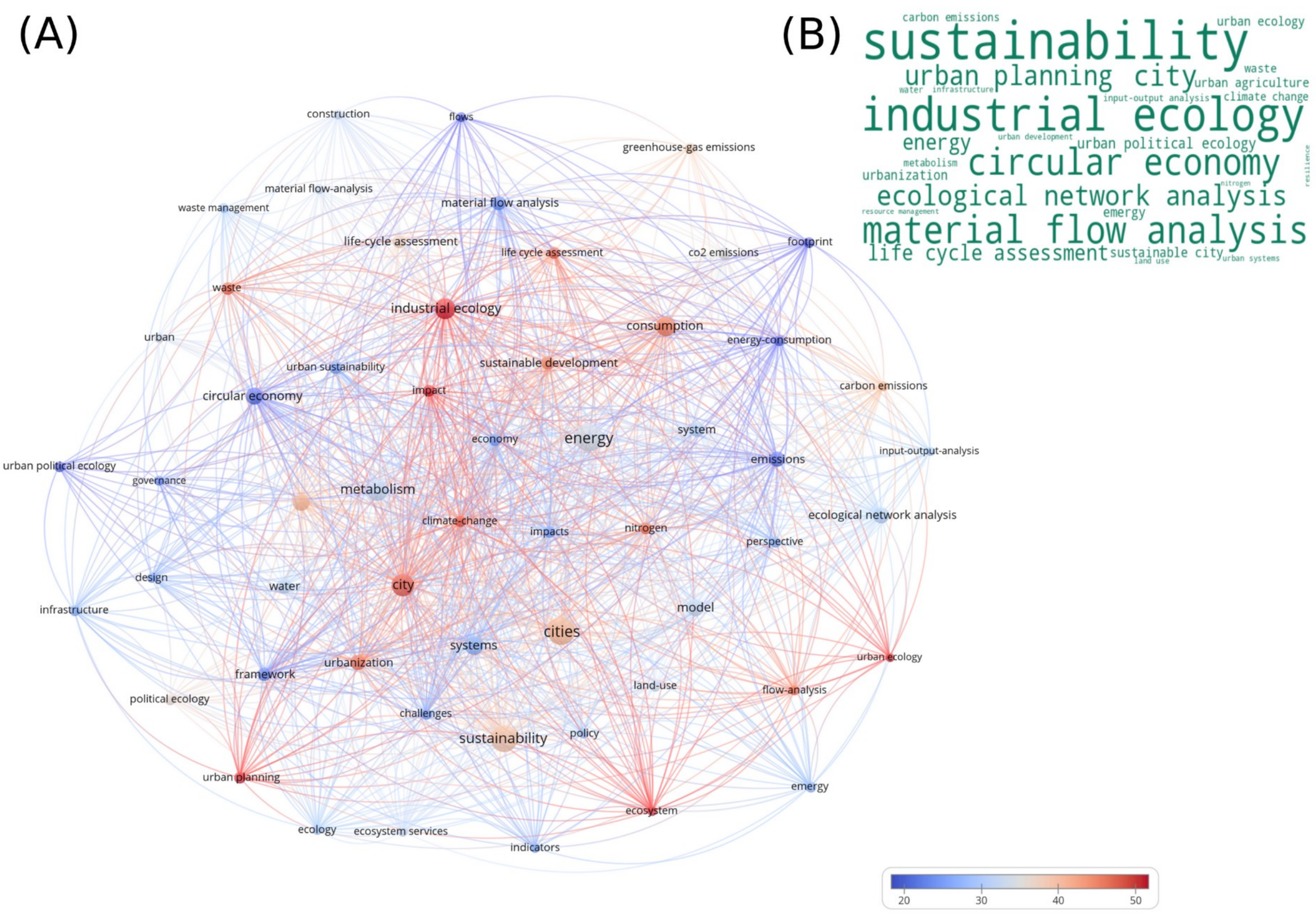

*Figure 2: A) Network of keywords of research articles analyzed. Nodes represent keywords appearing 10 or more times in the database. The color bar is the number of citations and node degree is scaled to the size of the node, representing the link's strength. B) Word cloud for the most frequent keywords in WoS* database associated to UM research.

*Table 1: Different visions of the relationship between Sustainability and UM summarized.*

| **How do articles or books reviewed address sustainability** | **How UM contributes to each conception of sustainability** | **Reference** |
|---|---|---|
| Sustainability minimizes resource extraction and pollutant emissions into the environment. | Quantifying urban metabolic flows helps minimize resource use and pollution. | (Newman, 1999) |
| Sustainability as a city's property that reduces its impacts on the environment, measured as ecological footprint. | Efficient urban metabolism reduces ecological footprint. | (Newman & Jennings, 2008) |
| A sustainable system conserves mass through recycling and is energy self-sufficient or is subsidized by sustainable inputs. | The study of UM is an integral part of the State of the Environment (SOE) reporting; thus, it provides measures of a city's sustainability. | (Kennedy *et al.*, 2011) |
| In the scaling theory, sustainability is the long-term balance between human development needs and the planet's environmental limits. | Scaling of the city's socioeconomic and biological entities lends to understand urban resource consumption or emissions efficiency. | (Fragkias *et al.*, 2013) |
| Sustainability goals include a decrease in resource utilization, thus reducing environmental impacts and improving human well-being. | Metabolic throughflows and efficiency indicators allow researchers to make effective suggestions to promote sustainable development. | (Zhang *et al.*, 2015) |
| Sustainability meets present needs without | UM provides parameters for evaluating the city's | (Céspedes Restrepo & |
| compromising future generations and encompasses ecological balance, economic stability, and social equity. | environmental impacts, enabling the design of sustainable urban planning policies. | Morales-Pinzón, 2018; Gavrilescu, 2011) |
| Sustainable practices require effective land use management, without undermining the Earth's long-term ability to provide ecosystem services. | The metabolic cycle informs waste management and tracks urban change. This metaphor is useful for drawing attention to how a city's metabolism changes over time. | (Pickett *et al.*, 2019) |
| Achieving a sustainable future requires cities to address the rise of poverty, hunger, resource consumption, and biodiversity loss. | Understanding urban scaling –as a way to approach UM– will help provide the foundations to develop effective urban sustainability policies. | (Cheng, 2023; National Academies of Sciences, Engineering, and Medicine, 2016) |
| Industrial or social metabolism theory frames material and energetic relationships between society and nature as sustainability challenges | UM analyses reveal unsustainable imbalances between demand and supply, and pollution flows. | (Fischer-Kowalski, 1998; Pincetl *et al.*, 2012) |

## 3.3. Distribution of Studies by Discipline, Spatial-Scale, and Methodology

Figure 3 shows the disciplines or research areas categorized according to the WoS, where Environmental Science & Ecology is the main field for UM research, followed by Green & Sustainable Science & Technology, and Environmental Engineering.

Within 320 studies in which we were able to identify the treated scale, 211 (65.9%) used the city as the unit of analysis, as shown in Figure 4.

We analyzed 369 unique articles from the WoS and selected 252 to identify the main methodologies used in UM, the scale at which it is applied. We also identified key articles and books to understand how sustainability is perceived in UM works. The methodologies reviewed here have diverse goals, use different data gathering approaches (top-down versus bottom-up), and are based on accounting or modeling. The 10 main methodologies are summarized in Table 2 and a more detailed description is shown in Supplementary Table 3. Material Flow Analysis (MFA), Life Cycle Assessment (LCA), Food-Energy-Water Nexus, Ecological Network Analysis (ENA), Input-Output Analysis (IO), and Emergy seem to be the preferred UM analysis methods (Figure 5A).

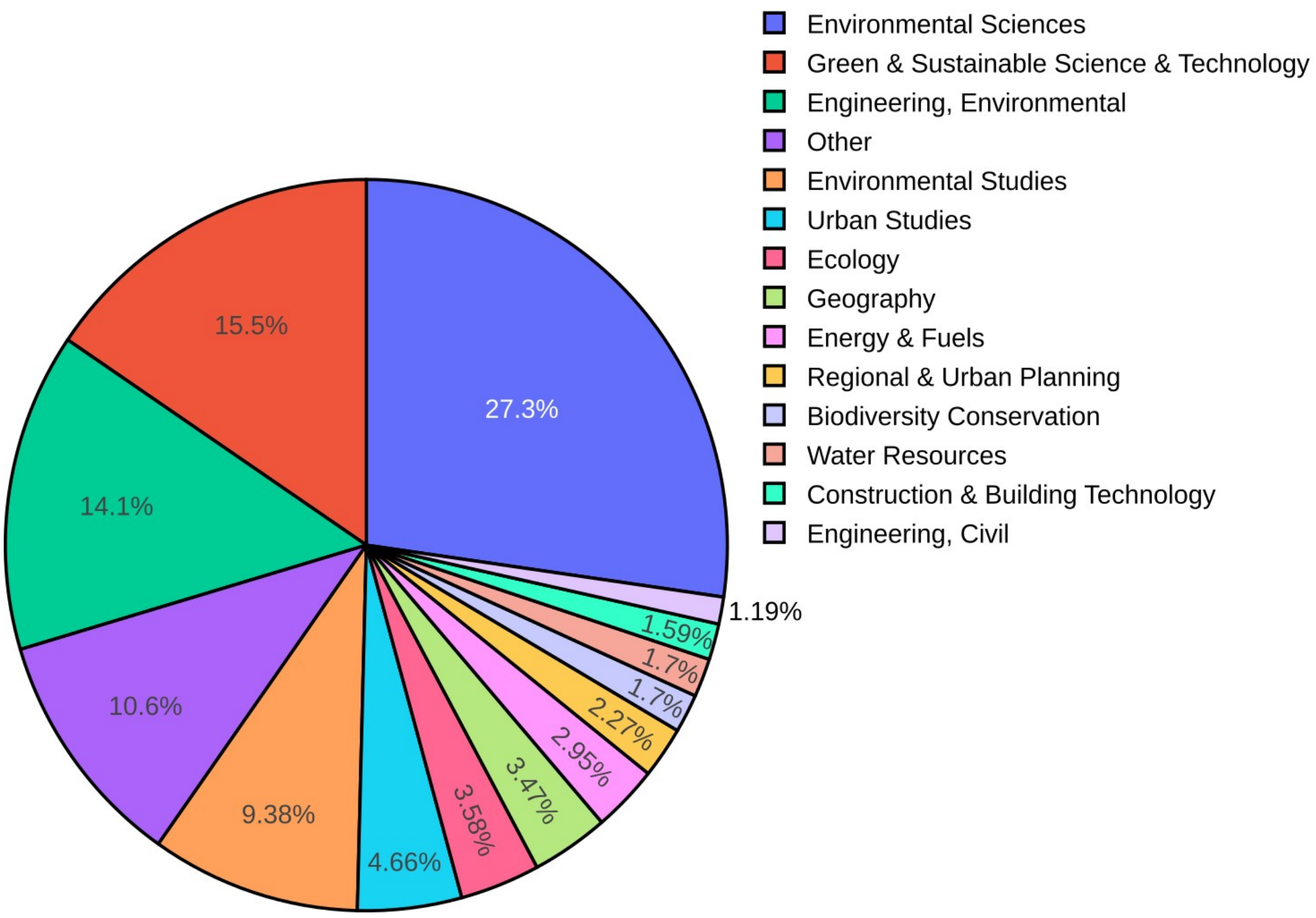

*Figure 3: UM and Sustainability research areas in the WoS.*

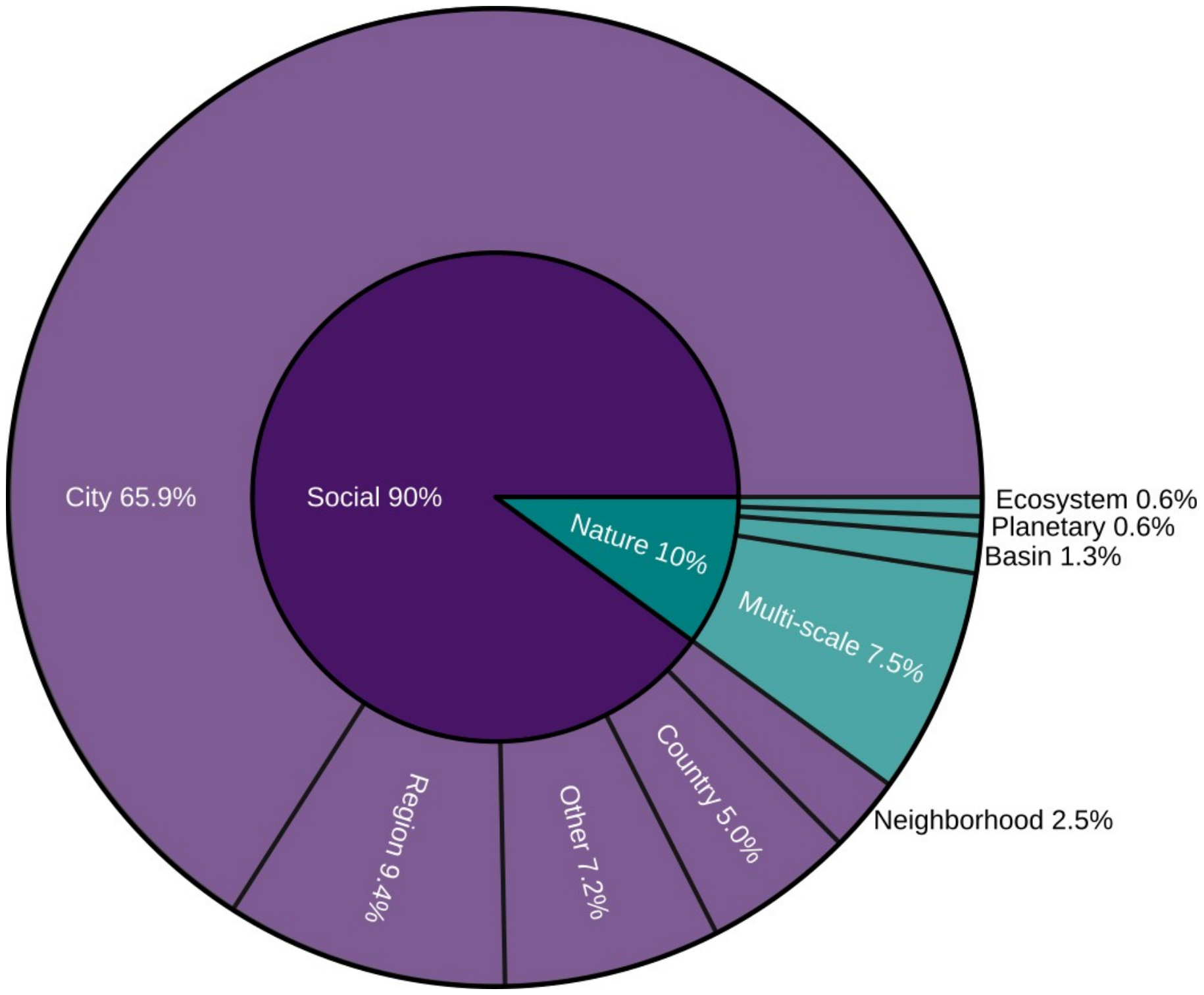

*Figure 4: The UM analysis spatial scale is shown as the relative frequency of articles in two realms in our literature review. Most articles (90%) are related to social definitions of the spatial extent in which these studies are conceived. The remaining 10% used nature-based definitions of the spatial scale to study UM.*

*Table 2: Main methods used to assess UM summarized. Type of Method (Stock or Flow) was determined through own research, while Type of Method (Accounting, Modeling, and process-based) was defined according to (Zhang et al., 2015).*

| **Method (Acronym)** | **Description** | **Goal** | **Input** | **Output** | **Type of Method *** | **Direction of Information** | **References** |
|---|---|---|---|---|---|---|---|
| Ecological Footprint (EF). or EF Assessment or Analysis (EFA) | Accounting for the biologically productive land and sea area required for resource consumption. | To compare resource use with land regeneration capacity, measured as land occupation in "global hectares" (gha). | Material (water [L], domestic waste [kg], food [kg]) and energy [kWh] demand. | Ecological footprint [gha] or other footprints such as carbon footprint [kg $CO_2$e] | Stocks<br><br>Accounting method | Both | (L. Huang *et al.*, 2015; Luck *et al.*, 2001; Rees & Wackernagel, 1996; Wackernagel *et al.*, 2006) |
| Ecological Network Analysis (ENA) | Quantitative simulation of structural and functional relationships within an ecosystem. | To analyze the interactions between ecosystem structure and function. | Input-output tables (see below). | Network representation of the city and its interactions. | Flow<br><br>Modeling method | Top-down | (Borrett *et al.*, 2018; Hannon, 1973; Zhang, 2013; Zhang *et al.*, 2009, 2015) |
| Emergy Analysis | Total energy used directly or indirectly for system functioning, expressed in solar emjoules. | To understand linkages among urban system. | Material (water [L], domestic waste [kg], food [kg]), energy [kWh], and monetary demand. | Flows aggregated in emergy units: the solar emjoule [sej]. | Flow<br><br>Non-material accounting method | Both | (Odum, 1996; Tan *et al.*, 2018; Y. Zhang *et al.*, 2009, 2015) |
| Energy-Water (EW), Food-Energy-Water (FEW), Nexus or similar | Analysis of consumption flows across food, energy, and water sectors. | To identify trade-offs and synergies within sectors. | Food [kg], energy [kWh], and water [L] demand. | Flow quantification and optimization models. | Flow<br><br>Accounting method or process-based method (Meng *et al.*, 2023) | Both | (Fan *et al.*, 2019; Meng *et al.*, 2023; Newell *et al.*, 2019; Peña-Torres *et al.*, 2022; Zhang *et al.*, 2020; Zhang *et al.*, 2015) |
| Exergy Analysis | Measurement of the maximum available energy for system equilibrium, indicating environmental impacts. | To provide an integrated analysis across diverse measurements. | Material (water [L], domestic waste [kg], food [kg]) and energy [kWh] demand. | Flows measured in exergy units [kJ/$m^3$]. | Stock<br><br>Non-material accounting method | Both | (Balocco *et al.*, 2004; Jorgensen & Mejer, 1977; Zhang, 2013) |
| Indicators (not explicit) | Urban metabolic flows (e.g., water and energy use) used as the city's sustainability indicators. | To establish a quantifiable value of urban sustainability. | Material (water [L], domestic waste [kg], food [kg]), and energy [kWh] demand. | Per capita water use, energy use, and solid waste generation. | Flow<br><br>Accounting | Both | (Newman, 1999) |
| Input-Output Tables (IO) or Analysis (IOA). Or Economic IO (EIO) and Environmental IO Analysis (EIOA) | Building tables that combine monetary and environmental flow analyses. | To use the IO analysis for environmental impact indicators, helping to refine the understanding of urban metabolic process stakeholders. | Monetary input-output tables (Leontief, 1986). | Sectoral distribution of resource consumption and impacts. | Flow<br><br>Modeling method | Top-down | (Chen & Chen, 2011; Fan *et al.*, 2019; Lombardi *et al.*, 2017; Zhang, 2013) |
| Life Cycle Assessment (LCA) | Evaluation of environmental burdens associated with products or activities. | To facilitate comparisons and identify supply chain hotspots. | Material (water [L], domestic waste [kg], food [kg]) and energy [kWh] demand. Land use [ha]. Industrial, and transport production (mass, volume, distance). | Inventory and impact assessment classified by LCA categories e.g., global warming potential (as carbon footprint) [kg$CO_2$e]. | Flow<br><br>Accounting method | Bottom-up | (Fan *et al.*, 2019; Newell *et al.*, 2019; SETAC, 1993; Zhang *et al.*, 2015) |
| Material-Flow Analysis (MFA) or Material-Energy-Flow | Study of physical material flows within a system based on mass balancing. Applied to specific | To analyze the relationships among material flows and environmental changes. | Material (water [L], domestic waste [kg], food [kg]), and energy | Flow matrices, data inventory for other methodologies | Flow<br><br>Accounting method | Both | (Fischer-Kowalski, 1998; García-Guaita *et al.*, |

| Method (Acronym) | Description | Goal | Input | Output | Type of Method * | Direction of Information | References |
|---|---|---|---|---|---|---|---|
| Analysis (MEFA). | elements or compounds is Substance-Flow Analysis (SFA). | | [kWh] demand. | (as LCA). | | | 2018; OECD, 2008; Y. Zhang, 2013) |
| Scaling | An analytical framework to characterize property variations based on system sizes. | To understand how size affects system properties. | Scaling metrics such as population and GDP. | Parameters describing scaling law function (scaling exponent). | Stock | Top-down | (Bettencourt, 2013; Bettencourt *et al.*, 2020; Ribeiro & Netto, 2024) |

## 3.4. Interactions Between Scale of Analysis and Methods

The methodologies often used at different spatial scales were also analyzed (Figure 5B). According to the data, all of the methodologies are used at the city level (which is the most studied level), with MFA, ENA, Nexus, IO, and LCA as the most frequent methods. This analysis also shows interesting gaps, including that the scaling analysis has not been used at the country level.

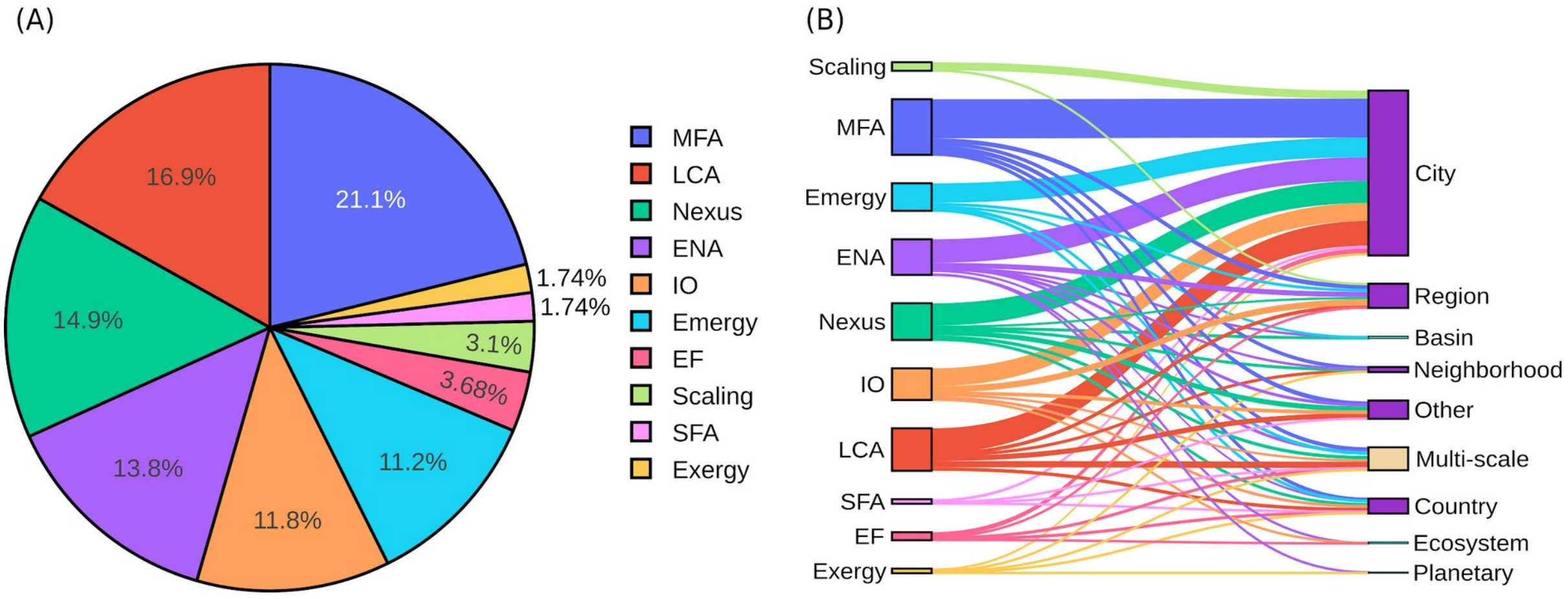

*Figure 5: A) Number of articles on UM and main methodologies used. Percentages are computed out of 516 articles in our dataset (369 unique articles). B) Interconnections between UM methods (left-hand side) and analysis scale (right-hand side) in the works on UM reviewed. The link sizes represent the number of articles, books, etc. from each method applied to each scale.*

## 3.5. The Diversity of Researchers in the UM and S Field

The countries with the largest contributions in terms of published articles are China and the USA, both accounting for more than 40% of the total database (Figure 6 A).

Figure 6B shows female participation in the WoS database. It displays the gender distribution of all authors, first author and last author.

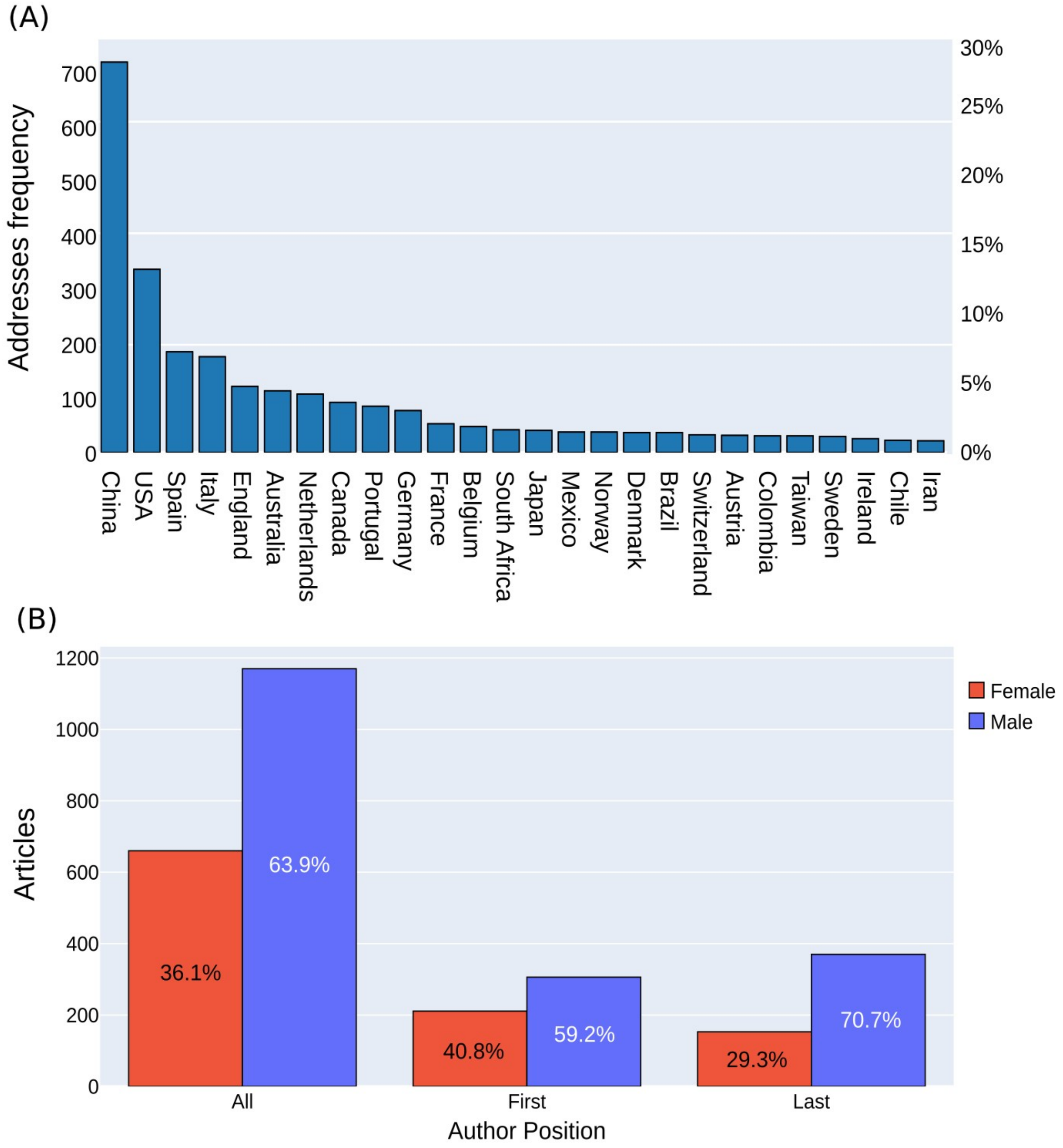

*Figure 6: A) Address frequency (based on all of the article's authors) per country for Urban Metabolism and Sustainability researches. Countries with less than two publications were excluded from this chart. B) Gender distribution of all authors whose genders were identified is shown (62.7% of the 2919 authors in the WoS database), for the first authors (65.4% with identified gender), and for the last authors (66.1% with identified gender).*

# 4. Discussion

We showed that Urban Metabolism (UM) is an emerging field characterized by significant growth in recent years. One of the possible explanations is that, after COVID-19, many scientists from different disciplines devoted their efforts to study COVID-19-related problems, and Urban Science might have not been the exception.

The multidisciplinary nature of UM is evident from the broad distribution of articles across various fields of knowledge. Notably, disciplines such as ecology, earth sciences, and engineering are prominently represented. While urban studies and planning –fields demonstrably germane to the study of cities– are also present, their contribution amounts to less than 7% of the overall research body. This observation suggests that a significant share of UM research stems from investigators with diverse disciplinary backgrounds, rather than solely from experts within the urban science realm.

The geographic distribution of UM research is concentrated primarily in China and the United States. It is important to note that this analysis focuses on the authors' country of origin , rather than on the study's specific location. For example, Hammelmann (2022) examined urban agriculture in Rosario, Argentina, but the research was documented in the USA due to the author's affiliation. When analyzing female participation in UM research (36.1%), we see that more female authors in this field than in overall scientific research (i.e. 26.8%, as reported by Huang *et al.*, 2020).

Interestingly, female participation increases in first authorship (reaching 40.8%) but decreases in last authorship (29.3%). Although this is not a general rule, usually the first author is the student or researcher who has undertaken the research work while the last author is the senior author who provides the intellectual input for the study, i.e. the group leader. This reveals that, although there seems to be fewer gender gaps in UM than in other research fields, this gap would apparently be greater for senior authors. This is relevant due to the fact that it has been well-established that diversity enhances scientific research (Oreskes, 2019), and that such diversity can be reflected in researchers' varying cultural origins and genders. Women constitute a minority of researchers globally (UNESCO, 2020), a phenomenon partly attributable to gender biases in academic evaluations and significant access biases to STEM fields by young researchers (Heilman, 2012; Huang *et al.*, 2020).

The distribution of methodologies employed in the literature reviewed is relatively homogeneous, with no single predominant method (Figure 5A). This suggests a lack of consensus regarding the optimal approach to assess UM or urban sustainability through a metabolic lens. Additionally, some studies combine multiple methods, such as IO for ENA (Tan *et al.*, 2018) or integrating the Food-Energy-Water Nexus with LCA (Meng *et al.*, 2023). Method integration and the scale of analysis show that most of the UM-related articles work at the city scale, without a preferred methodology at that level (Figure 5B). Articles using less common methodologies, such as EF and Exergy, seem to be mostly employed beyond the city scale. Few articles address natural scales, especially some of the most common methodologies, such as MFA, LCA.

The methodologies associated with the stock of resources are more uniformly distributed within Social and Natural scales compared to methodologies associated with flows. Five of the six most widely used methods rely on quantifying material and/or energy flows associated with urban functions, and a few methods delve into resource availability in natural systems (stocks) beyond quantification. Similarly, most methods are tailored to discuss urban systems in detail and lack the possibility, or the emphasis, to describe general relationships across urban systems. Scaling analysis offers a potential approach to address both of these limitations, although its prevalence in the literature is relatively low. This may be due to the fact that scaling research often fails to explicitly adopt an UM framework. Given the importance of understanding system dynamics for evaluating sustainability, promoting methodologies that can track stocks and find general relationships that can be applied to urban centers could be particularly important for global sustainability assessments, as proposed by the Earth Systems Boundaries framework (Bai *et al.* 2024) .

Our research shows that authors claim different visions of sustainability. While some conceptions are focused on resource consumption and waste disposal (efficiency), others delve more into holistic aspects of human wellbeing and ecosystem resilience, and do not restrict the analysis to flows within the city but incorporate a systemic coupling. A more general definition of sustainability could be relevant in order to highlight general first principles. Sometimes less is more.

Our analysis evidenced many challenges in UM and Sustainability. For example, understanding why after 2019 the field has experienced a decline in the number of published articles, and which methodologies have not been used at higher spatial scale, such as the national level. Given that countries function as distinct sociopolitical entities with unique environmental policies that significantly influence material and waste flows, investigating these dynamics at this scale offers a valuable avenue for research. This approach is particularly relevant considering the historical emphasis on national-level public policy implementation compared to city-scale interventions. Addressing socio-environmental problems at different scales is essential for global sustainability. Some UM methodologies intrinsically follow this perspective (e.g., longitudinal analysis in urban scaling, or telecoupling in accounting methods). However, there are still gaps in these approaches.

# 5. Conclusions

Urbanization, driven by increasing access to opportunities and resources, is a significant global change driver. As a complex socio-ecological system, cities offer a compelling subject for sustainable development research. Urban Metabolism (UM) provides a valuable interdisciplinary lens for analyzing urban environments from a sustainability perspective. By integrating frameworks from socio-ecological systems (Liu *et al.*, 2020; Ostrom, 2009), environmental impact assessment (Kennedy *et al.*, 2015; Pincetl *et al.*, 2012), and circular economy (Prendeville *et al.*, 2018), UM can offer a comprehensive understanding of urban sustainability. Our findings reveal that UM is an emerging field with a strong connection to sustainability and a relatively low gender bias compared to other disciplines. The field exhibited a decline in the number of articles published after 2019 and has a preference for quantifying metabolic flows over dynamic or structural system analysis. Furthermore, the concept of sustainability is interpreted variably within the literature, with a common emphasis on resource efficiency and waste reduction.

Key questions remain regarding the diverse conceptualizations of sustainability and metabolism. For instance, what are the benefits of viewing cities as organisms or ecosystems? What constitutes urban metabolism, and which processes are involved? Are there social evolutionary forces shaping this metabolism? The absence of a temporal dimension in UM research limits its ability to fully address sustainability challenges. Incorporating time-series analysis, transdisciplinary methodologies, and systems thinking is essential for planning sustainable cities in the face of climate and ecological crises.

# Declaration of Generative AI and AI-Assisted Technologies in the Writing Process

During the preparation of this work the author(s) used Chat GPT in order to rephrase sentences aiming at improving the text’s readability and to generate Python code for data analysis and visualizations. After using this tool/service, the author(s) reviewed and edited the content as needed and take(s) full responsibility for the publication.

# CRediT Authorship Contribution Statement

Mariana Bruning-González: Conceptualization, Data curation, Formal analysis, Investigation, Project administration, Visualization, Writing – original draft.
Horacio Samaniego: Conceptualization, Funding, Resources, Validation, Writing-review and editing.
José Ignacio Arroyo and Pablo Marquet: Validation, Writing-review and editing.

# Declaration of Competing Interest

The authors declare that they have no known competing financial interests or personal relationships that could have appeared to influence the work reported in this paper.

# Acknowledgments

This research was funded by the Chilean Agency for Research and Development (ANID) through Doctoral Grant # 21210418 to MB-G and through the Regular Fondecyt grants # 1211490 to HS. Additional funding was provided by a InES "Conocimiento + Género UACh" / INGE220001 / ANID grant. We thank Dr. Felipe Díaz-Alvarado for his insightful discussions and support. We would like to thank the authors of the illustrations used in the graphical abstract by Jason C. Fisher and Tracey Saxby, from IAN (ian.umces.edu/media-library) and Victoria Cataldo, chilean illustrator.

# Appendix. Supplementary Material

Download: Download Word document

# Data Availability

Data will be made available on request.